\documentclass[aps,amsfonts,prl,twocolumn,showpacs]{revtex4}
\usepackage{epsfig,amsmath,amssymb,bm,epsf,graphics,psfrag,verbatim}

\newcommand{\be}{\begin{equation}}
\newcommand{\ee}{\end{equation}}
\newcommand{\bea}{\begin{eqnarray}}
\newcommand{\eea}{\end{eqnarray}}

\begin{document}
\title{Local superfluid densities probed via current-induced superconducting phase gradients}
\author{David S. Hopkins}
\affiliation{Department of Physics and Frederick Seitz Materials
Research Laboratory, University of Illinois at Urbana-Champaign, 1110
West Green Street, Urbana, Illinois 61801-3080, USA}

\author{David Pekker}
\affiliation{Department of Physics and Frederick Seitz Materials
Research Laboratory, University of Illinois at Urbana-Champaign, 1110
West Green Street, Urbana, Illinois 61801-3080, USA}

\author{Tzu-Chieh Wei}
\affiliation{Department of Physics and Frederick Seitz Materials
Research Laboratory, University of Illinois at Urbana-Champaign, 1110
West Green Street, Urbana, Illinois 61801-3080, USA}

\author{Paul M.~Goldbart}
\affiliation{Department of Physics and Frederick Seitz Materials
Research Laboratory, University of Illinois at Urbana-Champaign, 1110
West Green Street, Urbana, Illinois 61801-3080, USA}

\author{Alexey Bezryadin}
\affiliation{Department of Physics and Frederick Seitz Materials
Research Laboratory, University of Illinois at Urbana-Champaign, 1110
West Green Street, Urbana, Illinois 61801-3080, USA}

\date{July 10, 2007}

\begin{abstract}
%
We have developed a superconducting phase gradiometer consisting of two
parallel DNA-templated nanowires connecting two thin-film leads. We have
ramped the cross current flowing perpendicular to the nanowires, and observed
oscillations in the lead-to-lead resistance due to cross-current-induced
phase differences. By using this gradiometer we have measured the temperature
and magnetic field dependence of the superfluid density and observed an
amplification of phase gradients caused by elastic vortex displacements. We
examine our data in light of Miller-Bardeen theory of dirty superconductors
and a microscale version of Campbell's model of field penetration.
\end{abstract}
\pacs{74.50.+r, 74.78.-w}
%
%
%
%
\maketitle

\noindent Gradients in the phase of the superconducting order parameter can be
generated by external magnetic fields and sensed using a superconducting
quantum interference device (SQUID), invented by Jaklevic et
al.~\cite{REF:Jaklevic1964}.
Applications of SQUIDS are numerous~\cite{SQUIDHB}. As demonstrated by Clarke
and others, SQUIDS may be used as magnetometers to detect remarkably small
magnetic fields, e.g.,  magnetic fields occuring in living
organisms~\cite{ClarkeSA}. A voltmeter based on the SQUID, also developed by
Clarke, is capable of measuring 10~fV,
i.e.~it is $10^5$ times more sensitive than a conventional semiconductor
voltmeter~\cite{SLUG}.
Micro- and nano-scale realizations of SQUIDs have been fabricated, e.g., in
 shunted Nb nanojunctions having a sub-micron hole~\cite{Lam} and, more
recently, using nano{\-}tubes~\cite{Grenoble}. SQUIDs have also been used to
study macroscopic quantum phenomena and devices,  such as magnetization
tunneling~\cite{Wernsdorfer97,Wernsdorfer99} and phase
qubits~\cite{Chiorescu03}.

In this Letter we report on a molecular-templated version of the Jaklevic et
al.~phase-sensing experiment, in which DNA-templated superconducting nanowires
are used instead of Josephson junctions to make a nanowire version of a SQUID
(i.e.~an N-SQUID).
Our main accomplishments are as follows.
(1)~By measuring the period of the resistance oscillations driven by an
externally injected supercurrent, we have performed a direct, local
measurement of the superfluid density and its dependence on temperature in
the vortex-free regime. Our results cannot be consistently described by the
phenomenological Gorter-Casimir and Ginzburg-Landau models. However, we do
find good agreement with Miller-Bardeen's dirty limit of the microscopic BCS
theory~\cite{REF:Bardeen1962}, which was previously examined for MoGe via a
different method by the Lemberger group~\cite{REF:LembergerPRL}.
(2)~We have observed and investigated the amplification of the current-induced
phase gradient that occurs when pinned vortices are present in the
cross-current-carrying lead. This amplification is brought about by a Lorentz
force, which acts on randomly pinned vortices and results in their reversible
displacement. In macroscopic settings, this physical phenomenon is the origin
of the Campbell law, i.e.~the dependence of the magnetic-field penetration
depth on the concentration of vortices~\cite{REF:Prozorov2003,
REF:Campbell1969,REF:Campbell1971,REF:Labusch1968}. Our measurements provide a
verification of the physics behind the Campbell law, but now at the
microscale, as well as the capability of obtaining the Labusch parameter,
i.e.~the average stiffness of the vortex pinning potential.

\begin{figure}
\includegraphics[width=6cm]{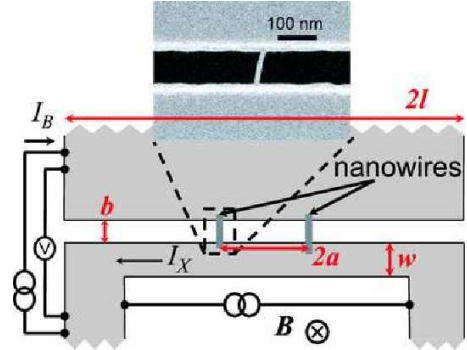}
\caption{(color online) Schematic of the DNA-templated two-nanowire device.
Two strands of DNA are stretched across a trench etched into \({\rm SiN/SiO}_2\) on
a \({\rm Si}\) chip.  The molecules and banks are sputter-coated with
\({\rm Mo}_{79}{\rm Ge}_{21}\).  A bias current \(I_\text{B}\) is applied through the
nanowires and a cross current \(I_\text{X}\) is applied through the horseshoe
lead, thus passing the contact points of the nanowires. A magnetic field \(B\)
is applied perpendicular to the plane of the leads and nanowires. One of the
two metal-coated DNA molecules is shown on the SEM micrograph. }
\label{FIG:device}
\end{figure}

Figure~\ref{FIG:device} shows the fabricated device.  First, two MoGe
nanowires are created by metal-coating a pair of suspended DNA
molecules~\cite{REF:Hopkins2005}. Then, a focused ion beam is used to cut one
of the leads into a horseshoe shape, and thus define a narrow strip through
which a cross current (\(I_\text{X}\)) can be applied.  Independently, a bias
current (\(I_\text{B}\)) can be applied through the nanowires, via which the
lead-to-lead resistance is measured~\cite{REF:Hopkins2007}. The typical width
of the cross-current strip is of the order of $700\,\text{nm}$ (e.g.,
$633\,\text{nm}$ for sample \# 2), which is  much smaller than the estimated
zero-temperature perpendicular magnetic field penetration depth
$\lambda_\perp = 46 \, \mu\text{m}$. Thus, the analyses of the results will
be carried out under the assumption that the cross current flows uniformly in
the cross-current strip. The width of the larger of the two leads,
$2l=17.33\,\mu\text{m}$ is smaller than $\lambda_\perp $, and thus any
external field will penetrate both leads without significant attenuation.
Moreover, we estimate that the effects of the magnetic field generated by the
cross current are negligible. The length of the nanowires, denoted by $b$, is
$100\,\text{nm}$ for sample \# 2, and the distance between them, denoted by
$2a$, which determines the spatial resolution of the device, is
$3.13\,\mu\text{m}$ for this sample. Devices with a better spatial resolution
(e.g. with $2a~265\,\text{nm}$) were demonstrated in a different
setup~\cite{REF:Hopkins2005}.

\begin{figure}
\includegraphics[width=7cm]{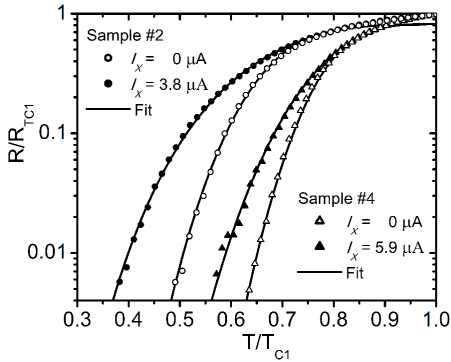}
\caption{Resistance vs.~temperature data for samples \#2 (circles) and \#4
(triangles).  The open shapes correspond to zero cross-current measurements
and the closed shapes correspond to a cross current that gives a resistance
close to a maximum (\(3.8\,\mu\text{A}\) for sample \#2 and
\(5.9\,\mu\text{A}\) for sample \#4).  The fits are obtained with the
following fit parameters:  sample \#2 -- \(R_\text{N} = 800 \, \Omega\),
\(I_{\text{c}1}(0) = 376\,\text{nA}\), \(I_{\text{c}2}(0) = 151\,\text{nA}\),
\(T_{\text{c}1} = 2.45\,\text{K}\), \(T_{\text{c}2} = 1.86\,\text{K}\); and
sample \#4 -- \(R_\text{N} = 700 \, \Omega\), \(I_{\text{c}1}(0) =
1066\,\text{nA}\), \(I_{\text{c}2}(0) = 317\,\text{nA}\), \(T_{\text{c}1} =
3.04\,\text{K}\), \(T_{\text{c}2} = 2.55\,\text{K}\)~\cite{footnote:Tc}. } \label{FIG:RT}
\end{figure}

At \(I_\text{X}=0\) we observe a broad resistive transition in the nanowires,
as the temperature $T$ is decreased. Moreover, as \(I_\text{X}\) is increased,
this transition periodically broadens and narrows back to its \(I_\text{X}=0\)
breadth.  We show this transition in Fig.~\ref{FIG:RT} for two samples, each
measured at two distinct values of \(I_\text{X}\) that correspond almost
exactly to the minimum and maximum observed transition breadths.  The
$T$ dependence of the resistance follows the theory of thermally
activated phase slips (TAPS)~\cite{REF:Little1967, REF:Langer1967,
REF:McCumber1970}, extended to the two-wire case~\cite{REF:Hopkins2005,
REF:Pekker2005}. This extended  theory involves a modified
free-energy barrier for phase slips, which accounts for the inter-wire
coupling mediated through the leads. In the short-wire limit,
the two-wire device has the current-phase relation of a single
Josephson junction with an effective critical current
\begin{align}
I_\text{eff}&=\frac{\sqrt{6}}{2}\frac{\hbar}{2 e}
\sqrt{(I_{\text{c}1}+I_{\text{c}2})^2 \cos^2\frac{\delta}{2}+
(I_{\text{c}1}-I_{\text{c}2})^2 \sin^2\frac{\delta}{2}}, \nonumber
\end{align}
tuned by the phase gain between the ends of the two nanowires \(\delta\) (see
Fig.~\ref{FIG:device}). For the cross-current experiment, \(\delta
= {\int}_{-a}^{a} \nabla \phi \cdot dr\), where $\phi$ is the phase of the
superconducting order parameter and the integral runs between the nanowire
ends via the horseshoe lead. We note that the oscillation period, which is the
main focus of this Letter, is independent of whether the nanowires are
operating in the short wire limit or not.
 Moreover,
\(I_{\text{c}1}\) and \(I_{\text{c}2}\) are the critical currents for the
nanowires, and are given by \(I_{\text{c}1,2} = I_{\text{c}1,2}(0)
[1-T/T_{\text{c}1,2}]^{3/2}\)~\cite{REF:Tinkham2002}. We obtain the device resistance \(R\) from the damped Josephson
junction formula~\cite{REF:Ivanchenko1968, REF:Ambegaokar1969} (Eq.~(9) in
Ref.~\cite{REF:Ambegaokar1969}), where we have used the normal-state
resistance \(R_\text{N}\) of the two parallel nanowires for the effective
shunt resistance of the junction.  Choosing \(\delta=0\) and \(\delta=\pi\),
we fit the lower and upper curves in Fig.~\ref{FIG:RT}.
\begin{figure}
\psfrag{R}{$R\,(\Omega)$}
\psfrag{I}{$I_\text{X}\,(\mu\text{A})$}
\includegraphics[width=7cm]{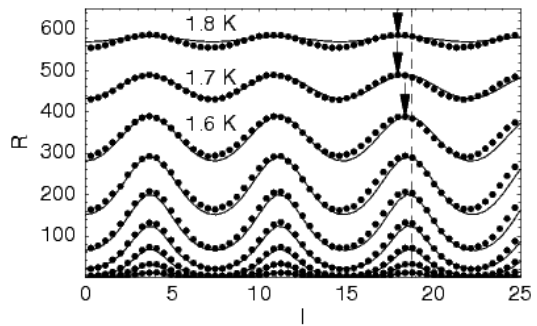}
\caption{Resistance vs.~cross-current data for sample \#2 at
temperatures ranging from \(1.0\,\text{K}\) to \(1.8\,\text{K}\) in
\(0.1\,\text{K}\) increments.
 The solid lines are fits using the same
 nanowire parameters as the fit for the \(R\)-\(T\) data in  Fig.~\ref{FIG:RT}
and for the period vs.~\(T\) data in Fig.~\ref{FIG:PT}.  The dashed vertical
line corresponds to the position of the peak at \(1\,\text{K}\), and is an aid
to see the decrease in period at higher temperatures.  The arrows point to
peaks and show the change in the  oscillation period with $T$.  }
\label{FIG:RIx}
\end{figure}
As shown in Fig.~\ref{FIG:RIx}, the resistance oscillates as a function of
\(I_\text{X}\), having a period on the order of \(7\,\mu\text{A}\) and an
amplitude that is  maximal in the middle of the resistive transition.  As the
arrows in Fig.~\ref{FIG:RIx} indicate, the period of the resistance
oscillation is temperature-dependent, in contrast with the case of
magnetic-field-induced oscillations, which appear to be temperature
independent~\cite{REF:Hopkins2005}. Thus, we have direct proof that the
oscillation is not due to a magnetic field induced by the cross current.

\begin{figure}
\psfrag{T}{$T/T_\text{c}$} \psfrag{Ns}{$n_s(T)/n_s(0)$} \psfrag{N2}{}
\vspace{-0.5cm}
\includegraphics[width=7cm]{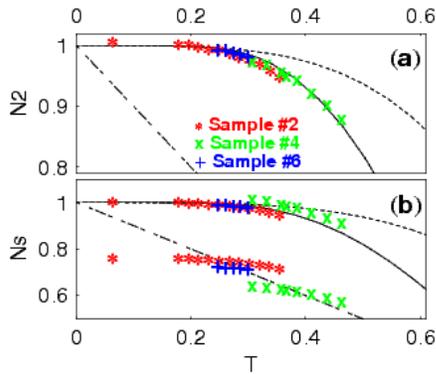}
\caption{(color online) Normalized superfluid density $n_s$
vs.~$T/T_\text{c}$. The solid line represents Miller-Bardeen's
  theory (Eq.~\ref{EQ:nsT}); the dotted line is
  Gorter-Casimir (GC) model; the dotted-dashed line is Ginzburg-Landau (GL) formula.
  Top: The $n_s(0)$ was chosen to optimize Miller-Bardeen's fit, the respective scaling
  parameters $\Delta I_\text{X}(0)$ being \(7.39\,\mu\text{A}\) (sample
  \#2),
  \(11.70\,\mu\text{A}\) (sample \#4), and
  \(7.30\,\mu\text{A}\) (sample \#6). The values of  $T_\text{c}$ used here, 5.64 K, 5.72 K, and 5.77 K
  for samples \#2, \#4,
and \#6, respectively, have been measured independently.
  Bottom: The same set of data is shown, but $n_s(0)$ was chosen to
  optimize the GC fit (top group of the points) or the GL fit (bottom
  group of points).
   }
\label{FIG:PT}
\end{figure}

The period of the resistance oscillation is determined by the condition that
the phase gain \(\delta\) be an integer multiple of \(2\pi\).
For the
case of a uniform cross current in a thin-film strip of width \(w\)
and thickness \(d\), \(\nabla \phi\) is related to \(I_\text{X}\)
through the superfluid density \(n_\text{s}\): \(\nabla \phi =
\delta/2 a = (I_\text{X}/n_\text{s})(2 m/w d e \hbar)\),
where \(m\) and \(e\) are the electron mass and charge.  Thus, by measuring
the period of the resistance oscillation vs. \(T\), we obtain the superfluid
density in the strip carrying the cross current via
\begin{align}
\Delta I_\text{X}(T)=\left(\frac{\pi wd}{a} \frac{\hbar e}{2 m}\right)
n_\text{s}=\Delta I_\text{X}(0) \frac{n_\text{s}(T)}{n_\text{s}(0)}.
\label{EQ:DIT}
\end{align}
The normalized period of the resistance oscillation (and hence the normalized
superfluid density) vs. $T$ is shown in Fig.~\ref{FIG:PT}.  To make the fits
we have used Miller-Bardeen's result, applicable to the dirty superconductor
case~\cite{REF:Bardeen1962},
\begin{align}
n_\text{s}(T)=n_\text{s}(0)\frac{\Delta(T/T_\text{c})}{\Delta(0)} \tanh
\left[ \frac{\Delta(T/T_\text{c})}{2 k_\text{B} T}\right],
\label{EQ:nsT}
\end{align}
where \(T_\text{c}\) is the critical temperature,  \(n_\text{s}(0)\) is the
zero-temperature superfluid density, and \(\Delta(T/T_\text{c})\) is the
universal BCS gap relation~\cite{REF:Tinkham1996}.  The fit to
Miller-Bardeen's formula was performed by allowing $n_s(0)$ (and hence
$\Delta I_\text{X}(0)$) to be an adjustable scaling parameter. It gives
better agreement than do the alternative theories, such as Gorter-Casimir and
Ginzburg-Landau. In fact, the fit to Miller-Bardeen's theory can be further
improved by slightly varying
 $T_\text{c}$. Almost perfect Miller-Bardeen-type fits are obtained with $T_\text{c}=5.22,
5.94,\,\text{and}\, 5.56\, \text{K}$ for samples \#2, \#4, and \#6,
respectively.
Combining the obtained fitting parameters \(\Delta I_\text{X}(0)\),
 \(I_{\text{c}1}\), \(T_{\text{c}1}\), \(I_{\text{c}2}\), and
\(T_{\text{c}2}\), we produce the theoretical curves in Fig.~\ref{FIG:RIx}
without any additional adjustable parameters.

 \begin{figure}
\includegraphics[width=7cm]{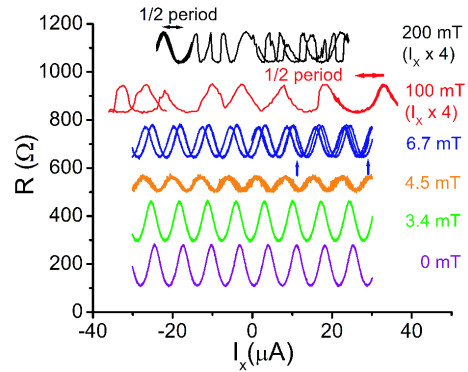}
 \caption{(color online) Resistance vs.~cross current measured at various values of
 the field at \(1\)~K.  For clarity, all the curves, except that of $0\,\text{mT}$,
 have been offset vertically by an integer multiple of $170\,\Omega$.
 The \(100\)~{mT} and \(200\)~mT curves have been
 horizontally magnified by a factor of \(4\), since the period
 became rather small at these fields. The curves show multiple random
 jumps at higer fields.  By measuring the
 resistance over a small range of cross currents several times,
  it was possible to measure at least half
a period of oscillation, as indicated by the horizontal arrows.  }
\label{FIG:RIxB}
\end{figure}

Having shown that our device is capable of sensing resistance
oscillations due to phase gradients created by external supercurrents,
we now turn our attention to the application of magnetic fields
\(\gtrsim H_{c1}\), which induce vortices in the thin-film leads and
thus create additional phase gradients. The penetration of vortices into
 mesoscopic samples was studied previously using other methods (see, e.g.
Refs.~\cite{REF:Martinis, REF:Baelus2005, REF:Arutunov}). All the measurements
presented in the following were performed on sample \#2.
Figure~\ref{FIG:RIxB} shows the resistance as a function of cross current at
several values of applied magnetic field.  At low fields, when vortices are
absent, the cross-current-induced oscillations are phase shifted with respect
to one another for distinct magnetic fields but with no detectable difference
in the oscillation period (see curves labeled \(0\)~mT and \(3.4\)~mT).  This
shift is due to the additional phase gradient caused by the Meissner currents
associated with the applied magnetic field.  At fields slightly above
\(H_{\text{c}1}\) for the wider lead (i.e. not the cross-current lead), it is
possible to see one of two types of resistance traces.  Sometimes, we observe
noisy oscillations with a smaller than expected amplitude (e.g. \(4.5\)~mT)
but with the same period as that at $B=0$.
  We believe that this is due to one or more vortices
that are rapidly wandering near one or both of the nanowires, i.e. a version
of motional narrowing.  At other times, we observe the type of oscillation as
observed below \(H_{\text{c}1}\), but with occasional jumps (\(6.7\)~mT) which
become more frequent when the field is larger.  At fields above
\(H_{\text{c}1}\) for the cross-current carrying lead (i.e.~the lead with the
larger \(H_{\text{c}1}\)), the period begins to decrease, as shown by the
highlighted segments for the \(100\)~mT and \(200\)~mT traces in
Fig.~\ref{FIG:RIxB}.  As vortex jumps are prevalent at these fields, and our
goal is to measure the period for a given vortex configuration, we determine
the period by measuring it within ``quiet regions'' (i.e. current intervals
exhibiting no jumps).  Examples of such regions are indicated
 in Fig.~\ref{FIG:RIxB} by horizontal arrows for the \(100\) and \(200\)~mT
traces.

\begin{figure}
\includegraphics[width=7cm]{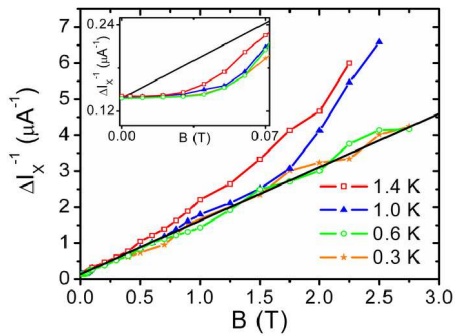}
\caption{(color online) Inverse of the oscillation period  vs.~field. The black line is  from  Campbell's formula with
\(\lambda_\perp=46\,\mu\text{m}\) and \(k=370\,\text{N}/\text{m}^2\). The
inset shows the low-field data.
}
\label{FIG:PB}
\end{figure}

We have investigated the resistance oscillation period  as a function of
magnetic field at four temperatures by making several scans of \(I_\text{X}\)
in ``quiet regions,'' in which it was possible to observe at least one peak
and one valley several times before a vortex jump occurred. At temperatures
below \(\sim 700\)~mK, at which the device has an undetectably small zero-bias
resistance, we obtained the period from large bias-current resistance
measurements (i.e.~beyond the linear response regime).  At each
 field and temperature, several measurements of the period were
recorded and averaged. Smoothed curves of the inverse period, obtained via a
moving average, are shown as functions of magnetic field in Fig.~\ref{FIG:PB}.
We observe a roughly linear increase of the inverse period; compared to its
zero-field value, the period at \(B=3\)~T is some thirty times smaller.  Such
behavior is analogous to the linear dependence of the perpendicular
penetration depth \(\lambda_\perp\) on applied magnetic field (in the regime
in which the vortex density is proportional to the field), i.e.~the Campbell
formula~\cite{REF:Prozorov2003, REF:Campbell1969, REF:Campbell1971}, adapted
for thin films: \( \lambda_{\text{C},\perp}=\big(\lambda^2+\big(\Phi_0 B/\mu_0
k(T)\big)\big)/d\). Here, \(k(T)\) is the vortex pinning force constant
(i.e.~the Labusch parameter~\cite{REF:Labusch1968}).
In the present setting, the phase fields from the vortices, displaced due to
\(I_\text{X}\), enhance the phase gradients resulting from the cross current,
 resulting in a  period for resistance oscillations that shortens with
\(B\).  We can therefore tie the Campbell prediction to the measured
period by assuming the usual relation \(n_\text{s}=m/\mu_0 e^2 d
\lambda_{\text{C},\perp} \).  From this and Eq.~(\ref{EQ:DIT}), we expect that
the inverse period should increase linearly with magnetic field for our
thin-film leads, \( (\Delta I_\text{X})^{-1}=(2a/w d)\big(\mu_0
\lambda^2/\Phi_0+\big(B/k(T)\big)\big) \).
  We estimate that $k\approx 370\,\text{N}/\text{m}^2$; other materials have yielded,
  e.g.,  $>100\,\text{N}/\text{m}^2 $~\cite{REF:Prozorov2003} and
  $\sim10^5\,\text{N}/\text{m}^2$~\cite{PesetskiLemberger}. Furthermore,
  the inferred value of $\lambda_\perp(B=0)$ is consistent with an estimate made
  using bulk penetration depth and film thickness
  data~\cite{footnote:penetration}. As Fig.~\ref{FIG:PB} shows, the inverse period does indeed
increase linearly, except at higher temperatures and fields. In those regimes,
inter-vortex interactions become important and hence, $k$ can be effectively
reduced. At high temperatures, vortices explore a larger area and are less
sharply pinned. For fields below \(H_{c1}\sim0.02\,\text{T}\),  no vortices
exist in the horseshoe lead, and the period does not depend
on the field.
 The  deviation from  linear field-dependence observed slightly above $H_{c1}$
 (Fig.~\ref{FIG:PB} inset)
 arises because in this regime the vortex density grows sublinearly with the field.

Regarding the future applications of our N-SQUID gradiometer, we remark that it can be used not only to measure the
local superfluid density, e.g., in connection with the fate of the Josephson
effect in ultra-thin film systems~\cite{Hermele}, but also to shed light on
settings in which local fluctuations of the superfluid density have been
predicted, such as highly disordered s-wave 2D superconductors~\cite{Trivedi}
and superconductors having magnetic impurities~\cite{Simons}.

{\it Acknowledgments\/}--- This work was supported by DOE DEFG02-96ER45434
and DEFG02-91ER45439 and by NSF
DMR 0134770. DSH and AB  acknowledge the access to the fabrication facilities
at the Frederick Seitz Materials Research Laboratory.

\end{document}